\documentclass[preprint,12pt,authoryear]{elsarticle}

\usepackage[T1]{fontenc}
\usepackage[utf8]{inputenc}
\usepackage{lmodern}
\usepackage{amsmath,amssymb,amsthm,mathtools}
\usepackage{graphicx}
\usepackage{booktabs}
\usepackage{multirow}
\usepackage{array}
\usepackage{makecell}
\usepackage{xcolor}
\usepackage{colortbl}
\usepackage{algorithm}
\usepackage{algpseudocode}
\usepackage{hyperref}
\usepackage{url}
\usepackage{enumitem}
\usepackage{siunitx}
\usepackage{caption}
\usepackage{float}        
\usepackage{placeins}     

\usepackage{etoolbox}
\AtBeginEnvironment{tabular}{%
  \hyphenpenalty=50%
  \exhyphenpenalty=50%

}
\AtBeginDocument{%
  \setlength{\parskip}{0.5\baselineskip}%
}

\hypersetup{
  colorlinks=true,
  linkcolor=blue!50!black,
  citecolor=blue!50!black,
  urlcolor=blue!50!black,
  pdftitle={Parallelized Hierarchical Connectome (PHC)},
  pdfauthor={Po-Han Chiang}
}

\newcommand{\R}{\mathbb{R}}

\newcommand{\softplus}{\mathrm{softplus}}
\newcommand{\rmsnorm}{\mathrm{RMSNorm}}

\newcommand{\heaviside}{\Theta}
\newcommand{\Wsyn}{W_{\mathrm{syn}}}
\newcommand{\Wstruct}{W_{\mathrm{struct}}}
\newcommand{\Mtopo}{M_{\mathrm{topo}}}
\newcommand{\Min}{M_{\mathrm{in}}}

\newcommand{\Mhebb}{M_{\mathrm{hebb}}}
\newcommand{\Vmem}{V_{\mathrm{mem}}}

\newcommand{\Vthstat}{V_{\mathrm{th,stat}}}

\newcommand{\Nmax}{N_{\mathrm{max}}}
\newcommand{\Mstep}{M_{\mathrm{step}}}

\newcommand{\Isyn}{I_{\mathrm{syn}}}
\newcommand{\Lconv}{\mathcal{L}_{\mathrm{conv}}}
\newcommand{\Ltask}{\mathcal{L}_{\mathrm{task}}}
\newcommand{\Lrate}{\mathcal{L}_{\mathrm{rate}}}
\newcommand{\Lvolt}{\mathcal{L}_{\mathrm{volt}}}
\newcommand{\Ltotal}{\mathcal{L}}
\newcommand{\bigO}{\mathcal{O}}
\newcommand{\BigTheta}{\Theta}

\definecolor{tabgray}{gray}{0.90}

\journal{Neurocomputing}

\begin{document}

\begin{frontmatter}

\title{Parallelized Hierarchical Connectome: A Spatiotemporal Recurrent Framework for Spiking State-Space Models}

\author[1,2,3,4]{Po-Han Chiang\corref{cor1}}
\ead{phc@nycu.edu.tw}

\cortext[cor1]{Corresponding author.}

\affiliation[1]{organization={Interdisciplinary Master's Program in Brain Technology, College of Electrical and Computer Engineering, National Yang Ming Chiao Tung University},
                country={Taiwan (R.O.C.)}}
\affiliation[2]{organization={Institute of Intelligent Bioelectrical Engineering, College of Electrical and Computer Engineering, National Yang Ming Chiao Tung University},
                country={Taiwan (R.O.C.)}}
\affiliation[3]{organization={Department of Electronics and Electrical Engineering, College of Electrical and Computer Engineering, National Yang Ming Chiao Tung University},
                country={Taiwan (R.O.C.)}}
\affiliation[4]{organization={School of Medicine, College of Medicine, National Yang Ming Chiao Tung University},
                country={Taiwan (R.O.C.)}}

\begin{abstract}
This work presents the Parallelized Hierarchical Connectome (PHC), a general 
architectural framework that upgrades temporal-only State-Space Models (SSMs)
into spatiotemporal recurrent networks. Conventional SSMs achieve parallel-scan
training but are limited to temporal recurrence, lacking lateral or feedback
interactions within a single timestep. PHC maps the diagonal SSM core to a
shared Neuron Layer and inter-neuronal communication to a shared Synapse Layer
of hierarchical regions, reconnected by a Multi-Transmission Loop iterating
spatial recurrence within each temporal window, at parameter complexity
$\Theta(D^{2})$ versus $\Theta(D^{2}L)$ of stacked SSMs. This spatiotemporal
framework enables the seamless integration of neuro-physical priors
intractable for standard SSMs, including adaptive leaky integrate-and-fire
dynamics, synaptic delay, short-term plasticity, Dale's Law with
E/I-asymmetric topology, and spike-timing-dependent plasticity. The framework is instantiated as PHCSSM, the first spiking SSM
that integrates all five biological priors and is evaluated on long-sequence
data ($T = 405$ to $17{,}984$ on the UEA Multivariate Time-Series
Classification Archive), achieving test accuracy competitive with
state-of-the-art SSM baselines at 1{,}312 to 4{,}891 trainable parameters
(1 to 4 orders of magnitude smaller than every baseline). PHCSSM further
admits a sequential recurrent spiking neural network (RSNN) deployment mode
that converges asymptotically to the parallel-scan training mode without
ANN-to-SNN conversion, with cross-backend reproducibility verified across
four hardware backends
(x86 CPU, H100 GPU, Cortex-A76, Cortex-M4F) including end-to-end fp32 deployment
on the Cortex-M4F microcontroller (40~KB SRAM, 128~KB Flash). PHCSSM thereby
bridges parallel-scan SSM and biologically grounded RSNN, two paradigms with
previously incompatible training regimes, into a single architecture and
trained weights.
\end{abstract}


\end{frontmatter}

\section{Introduction}
\label{sec:intro}

Linear State-Space Models (SSMs) such as S4 and Mamba have advanced sequence modeling by combining the expressive power of recurrent networks with the parallel-training efficiency of Transformers \citep{Gu2022a,Hasani2022,Smith2023,Orvieto2023,RuschRus2025}. Through a linear time-invariant formulation, these models leverage associative parallel scans to achieve $\bigO(\log L)$ training complexity on sequences of tens of thousands of timesteps. Successive GPU generations have compounded this advantage: the Nvidia V100 to A100 to H100 transition raised aggregate tensor throughput and memory bandwidth by roughly an order of magnitude, and parallel-scan-friendly architectures absorb the bulk of this generational uplift while sequential primitives extract only a small fraction, so the architectures that scan in $\bigO(\log L)$ inherit the dominant share of the modern compute dividend. The efficiency, however, comes at a structural cost: to remain parallel-scan-compatible, modern SSMs constrain their state-transition matrices to diagonal form \citep{Gupta2022,Gonzalez2024,Farsang2025}, compressing cellular decay and inter-neuronal communication into a single per-neuron operator and decoupling neurons within each timestep (Figure~\ref{fig:fig1}A). Three primitives central to cortical computation are therefore inexpressible in standard SSMs: within-timestep lateral and feedback projections, sign-restricted excitatory and inhibitory weights, and state-dependent synaptic transmission. SSMs scale to long sequences but produce architecturally homogeneous internal dynamics far removed from biological neural circuits, foreclosing their use as substrates for digital-twin or neuroscience-grade circuit modelling.

\begin{center}
\includegraphics[width=\linewidth]{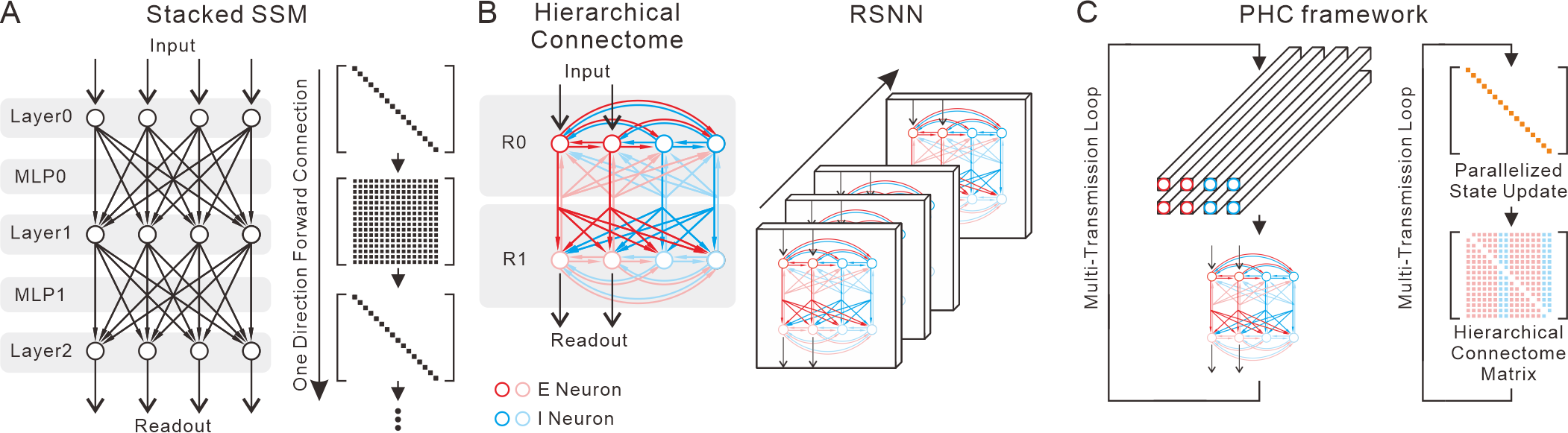}
\end{center}
\captionof{figure}{Comparison of stacked SSMs, recurrent spiking neural networks (RSNNs), and the Parallelized Hierarchical Connectome (PHC) framework. \textbf{(A) Stacked SSMs.} \emph{Left}, network diagram of $L$ diagonal-state-transition layers (Layer$_{0}$, Layer$_{1}$, Layer$_{2}$) interleaved with $L$ unidirectional MLPs (MLP$_{0}$, MLP$_{1}$), with input at top and readout at bottom. \emph{Right}, the corresponding stack of weight matrices alternating diagonal state-transition blocks and dense MLP blocks, illustrating the one-direction forward connection. The temporal scan is parallel but the spatial structure is strictly feedforward, with no lateral or feedback interactions within a layer. \textbf{(B) RSNNs.} \emph{Left}, single-timestep network diagram of hierarchical regions (R0, R1) with biologically constrained excitatory (red) and inhibitory (blue) populations and full lateral and feedback connectivity; \emph{right}, the same circuit unrolled across timesteps, illustrating that state propagation between timesteps is strictly sequential, foreclosing parallel-scan training. \textbf{(C) PHC framework.} \emph{Left}, the Multi-Transmission Loop (MTL) iterating spatial recurrence within each temporal window; \emph{right}, factorisation of the within-timestep operator into a diagonal parallel-scan core (Parallelized State Update) and a hierarchical connectome matrix ($\Wsyn \odot \Mtopo$) with biologically constrained block structure. Temporal dynamics are computed by a single shared $\bigO(\log T)$ parallel scan, while lateral and feedback interactions are handled within each timestep through the MTL, replacing depth-stacking with spatiotemporal recurrence at $\BigTheta(D^{2})$ parameter complexity.}
\label{fig:fig1}

Recurrent spiking neural networks (RSNNs) built from biological primitives possess precisely the dynamics that diagonal SSMs lack. Adaptive leaky integrate-and-fire (ALIF) dynamics endow each neuron with intrinsic temporal memory through slowly activating ionic currents that raise the firing threshold during sustained activity, mimicking cortical spike-frequency adaptation \citep{BendaHerz2003,Bellec2018}; short-term plasticity (STP) makes synaptic transmission state-dependent through presynaptic calcium-mediated facilitation and vesicle-pool depletion and recovery \citep{TsodyksMarkram1997,Mongillo2008}; Dale's Law enforces excitatory or inhibitory sign consistency at each presynaptic neuron, the structural basis of cortical E/I balance \citep{Markram2004,Dehghani2016}; within-timestep lateral and feedback projections support the hierarchical inter-areal interactions characteristic of primate cortex \citep{FellemanVanEssen1991}; spike-timing-dependent plasticity (STDP) provides a locally causal Hebbian learning channel grounded in NMDA-receptor-gated calcium dynamics \citep{BiPoo1998,FremauxGerstner2016}. Together these mechanisms make SNNs and biologically-constrained RNNs natural substrates for digital twins of cortical microcircuits and for computational-neuroscience investigations of how neural dynamics support behaviour \citep{Bellec2020,Yu2022}. The same recurrent dependencies that yield these dynamics, however, enforce strictly sequential execution: training cost scales linearly with sequence length, BPTT through spikes additionally requires surrogate-gradient approximations whose stability degrades with depth \citep{ZenkeGanguli2018,Bengio1994,Pascanu2013}, and RSNN benchmarks in the literature are largely confined to sequences of a few hundred to a few thousand timesteps. Modern accelerators amplify rather than alleviate this bottleneck. Transformer and SSM workloads convert each generational uplift in arithmetic throughput and memory bandwidth into proportionally faster training, but an RSNN's per-timestep dependency (Figure~\ref{fig:fig1}B) leaves the GPU underutilised regardless of available VRAM or tensor-core capacity; the compute dividend that has driven the SSM literature therefore bypasses RSNN architectures by construction and widens, rather than narrows, the practical training gap as hardware scales.

A separate line of work, ANN-to-SNN conversion \citep{Cao2015,Rueckauer2017,Sengupta2019,Bu2022}, sidesteps the native-SNN training cost by first training a conventional rate-coded ANN and then transferring its weights to a SNN that emulates the ANN's continuous activations via firing rates over a deployment-time emulation window; the resulting SNN inherits the parallel-training scalability of the source ANN but at the cost of a conversion-stage accuracy gap of typically 1 to 5 percentage points \citep{Sengupta2019}, a deployment-time emulation window of tens to hundreds of timesteps for spike-rate convergence, and the loss of the locally causal learning channels (in particular STDP) that natively trained spiking models provide. Liquid State Machines (LSMs) sidestep this training difficulty by freezing the recurrent spiking reservoir entirely and training only a linear readout \citep{Maass2002}, but the price is a randomly initialised, task-blind reservoir whose lateral connectivity cannot be adapted to input statistics. Recent spiking-SSM hybrids parallelise standard LIF dynamics \citep{StanRhodes2024,Zhong2024,Shen2025} but accommodate neither adaptive thresholds, multiplicative facilitation and recovery, sign-restricted Dale connectivity, nor within-recurrence lateral projections, leaving the simultaneous achievement of biological richness, end-to-end natively-trained spiking recurrent connectivity, and parallel scalability vacant in the current architectural landscape.

This architectural gap has direct consequences for two scientific--engineering frontiers. (A)~Edge biomedical computing (ECG and EEG monitoring, motor-imagery brain--computer interfaces, and closed-loop neuromodulation devices) requires sequence models that fit microcontroller-class memory budgets, run at physiological sampling rates on battery power, and adapt online to per-subject signal drift. Current SSMs deliver the temporal scaling but cannot supply the sign-restricted excitatory--inhibitory structure or the local Hebbian online learning needed for biologically-aligned closed-loop adaptation. On the other hand, current SNNs deliver the biological structure but cannot train at the multi-thousand-timestep scales of the underlying physiological signals. (B)~Computational neuroscience needs models that preserve the cellular and synaptic mechanisms whose computational role it seeks to study and yet train tractably on the long behavioural recordings produced by modern electrophysiology and chronic implants; here too the SSM-versus-SNN trade-off is decisive: neither side is usable as-is. Closing this gap is therefore not an architectural curiosity but a precondition for biologically-grounded sequence models to be trained on long real-world physiological signals and deployed on the resource-constrained hardware where such signals are acquired.

\begin{figure}[!htbp]
\centering
\includegraphics[width=\linewidth]{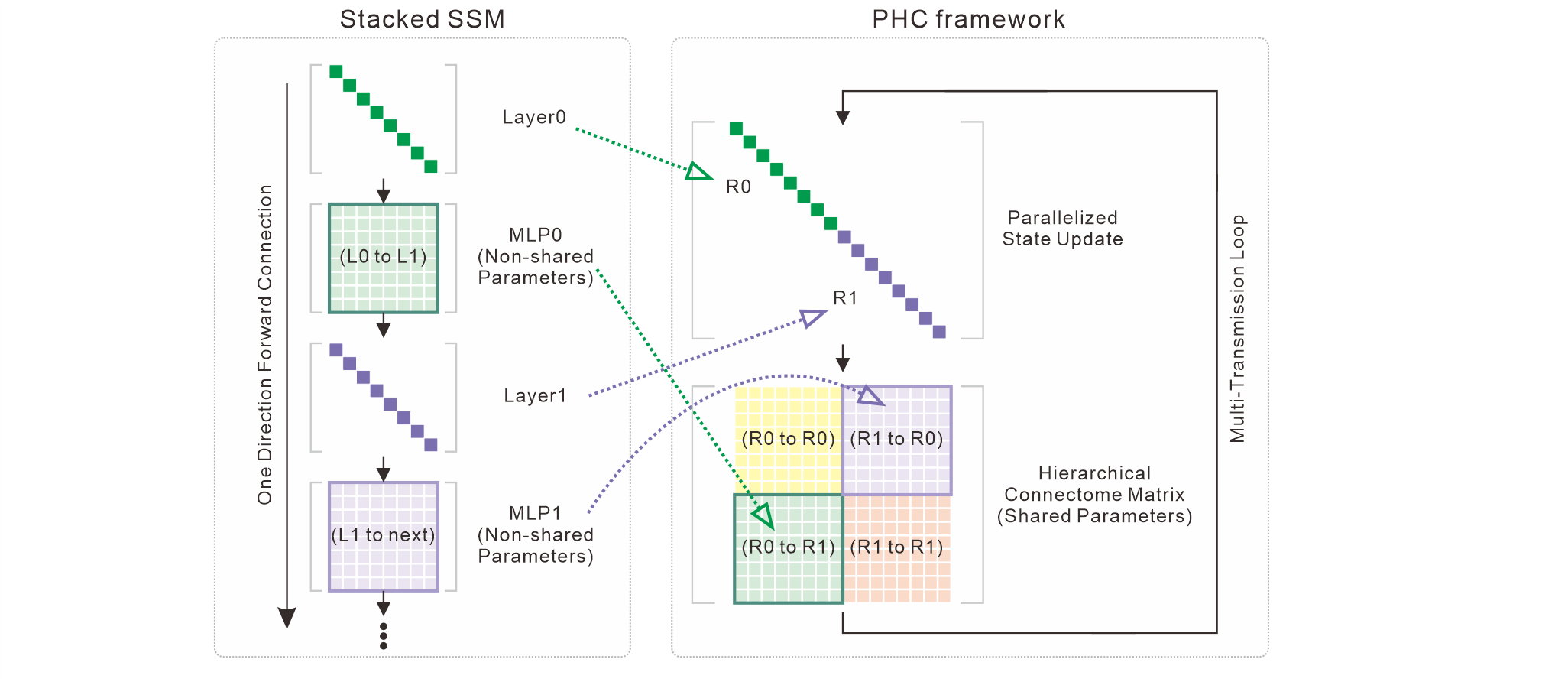}
\caption{Structural isomorphism between stacked SSMs and the PHC framework. \emph{Left}, a conventional $L$-layer stacked SSM, where $L$ independent diagonal state-transition matrices (Layer 0, Layer 1) are interleaved with $L$ independent dense MLPs (MLP$_{0}$, MLP$_{1}$), each with non-shared parameters, forming a unidirectional forward connection. \emph{Right}, the PHC framework collapses this vertical stack into a single spatial plane. Each diagonal layer maps to a region (R0, R1) within a shared Neuron Layer (NL) whose parameters are reused across all regions (Parallelized State Update). Each inter-layer MLP maps to a sub-block of the Hierarchical Connectome Matrix, which consolidates all inter-neuronal communication into a single shared Synapse Layer (SL) with biologically constrained connectivity (Dale's Law, topology mask, zero-diagonal). Unlike the stacked architecture's unidirectional forward path, the Connectome Matrix encodes bidirectional inter-region projections (e.g., R0~$\to$~R1 feedforward and R1~$\to$~R0 feedback) as well as intra-region lateral connections (R0~$\to$~R0, R1~$\to$~R1). The MTL iteratively circulates signals between the NL and the SL within each timestep, recovering the logical processing depth of $L$ stacked layers with only $\BigTheta(D^{2})$ shared parameters.}
\label{fig:fig2}
\end{figure}

To resolve this trade-off, this work introduces the Parallelized Hierarchical Connectome (PHC), an architectural framework that removes the connection-induced barrier to parallelisation (Figures~\ref{fig:fig1}C and~\ref{fig:fig2}). The PHC framework factorises the within-timestep operator into a diagonal core for per-neuron temporal evolution and a hierarchical connectome for inter-neuron communication, coupled through a within-timestep Multi-Transmission Loop (MTL); the temporal axis is integrated by an $\bigO(\log T)$ parallel scan even when the connectome carries learnable lateral and feedback weights. Demonstrated as a spiking SSM instantiation of this framework, PHCSSM simultaneously delivers two capabilities that have been mutually exclusive in prior parallel-scan architectures: (i)~parallel-scan training over the temporal axis, and (ii)~full integration of biological mechanisms: Dale's-Law-enforced E/I balance, ALIF dynamics, Tsodyks--Markram STP, and STDP.

Together, these two capabilities close the train--deploy loop that has historically split biologically grounded sequence modelling. PHCSSM trains as the parallel-scan PHC framework (Figures~\ref{fig:fig1}C and~\ref{fig:fig2}), restoring the GPU parallelism that the sequential per-timestep dependencies of conventional RSNN training cannot exploit; the same trained weights then redeploy as the conventional sequential RSNN (Figure~\ref{fig:fig1}B) for chip-class inference, preserving the biological constraints throughout. The closed-loop biomedical computing scenarios identified above (long physiological-signal sequence modelling on microcontroller-class hardware with online STDP-based adaptation) are thereby addressable by a single architecture without architectural compromise at either train time or deployment time.

This work makes four contributions. (1)~\emph{Intra-step spatiotemporal decoupling}: PHC is the first SSM framework introducing learnable lateral and feedback connection weights within the SSM recurrence while preserving $\bigO(\log T)$ parallel-scan training, decoupling temporal parallel scan from a spatial hierarchical connectome through the MTL. (2)~\emph{Parallelised neuro-physical dynamics}: log-domain affine-recurrence formulations of ALIF dynamics, Tsodyks--Markram short-term plasticity, and exponentially decaying spike-timing-dependent plasticity eligibility traces enable parallel prefix-sum training of biologically realistic non-linear recurrences without sacrificing scalability. To the best of the author's knowledge, PHCSSM is the first spiking model evaluated across the full six-dataset UEA-MTSCA benchmark under five simultaneous neuro-physical constraints (ALIF, synaptic delay, Tsodyks--Markram STP, Dale's Law with E/I-asymmetric topology, and STDP). (3)~\emph{Native online learning}: a STDP module exploits PHCSSM's genuine binary spike representation to provide a locally causal Hebbian learning signal complementary to backpropagation, a capability structurally unavailable to continuous-valued SSMs and to rate-coded ANN-to-SNN conversion pipelines. (4)~\emph{Cross-backend train--deploy verification}: the same trained weights produce bit-identical RSNN-mode predictions across four hardware backends (x86 CPU, H100 GPU, Cortex-A76, and Cortex-M4F), enabling end-to-end deployment on the Cortex-M4F microcontroller (40~KB SRAM, 128~KB Flash) at 1{,}312 to 4{,}891 trainable parameters, 7- to 35-fold CPU-deployment speedup, and microsecond-scale per-timestep streaming inference on commodity edge hardware.

\section{Related Work}
\label{sec:related}

\paragraph{Diagonal SSMs.} Modern SSMs achieve $\bigO(\log L)$ parallel training by enforcing diagonal state-transition matrices. The lineage from S4 \citep{Gu2022a} through S4D \citep{Gu2022b}, S5 \citep{Smith2023}, S7 \citep{Soydan2024}, and LRU \citep{Orvieto2023} progressively demonstrated that diagonal restriction preserves modeling capacity. Mamba \citep{GuDao2023} adds input-dependent state transitions for selective filtering. LinOSS \citep{RuschRus2025} employs second-order harmonic oscillator dynamics with block-diagonal transitions. LrcSSM \citep{Farsang2025} imposes diagonal Jacobian constraints on liquid-resistance liquid-capacitance networks. Across this family no model introduces within-timestep lateral interaction; representational depth comes through stacking independent blocks at $\BigTheta(D^2 L)$ parameter cost.

\paragraph{Spiking SSMs.} Recent spiking SSMs adapt diagonal SSM cores to discrete spike communication. Binary-S4D \citep{StanRhodes2024} applies binary activation to S4D states. SPikE-SSM \citep{Zhong2024} decomposes membrane potential for parallel computation. SpikingSSM \citep{Shen2025} approximates LIF dynamics through a surrogate dynamic network. These approaches focus on parallelizing standard LIF dynamics but treat inter-neuronal connections as unconstrained dense matrices, omit Dale's Law and short-term plasticity, and rely solely on surrogate gradients without timing-dependent online learning.

\paragraph{Lateral connections.} Several attempts introduce spatial structure into parallelizable sequence models. The Permutation-and-Diagonal SSM \citep{Terzic2025} factorizes the state transition matrix as $A(u) = P(u)\cdot D(u)$, preserving parallel-scan complexity through one-to-one permutation routing without learnable weighted lateral connections. xLSTM \citep{Beck2024} provides two cell designs: sLSTM with learnable lateral connections (non-parallelizable) and mLSTM (parallelizable but lateral-free). GraphS4mer \citep{Tang2023} and Graph Mamba \citep{BehrouzHashemi2024} confine spatial interaction to pre- or post-recurrence aggregation external to the SSM core. None of these attempts achieves within-timestep iterated weighted lateral interaction within a parallelizable recurrence: this is the gap PHC addresses.

\paragraph{Biological priors and online learning.} Biological priors have been incorporated piecewise into recurrent models. The liquid neural network family \citep{Hasani2021,Farsang2024a,Farsang2024b} introduces state-dependent membrane time constants and capacitance. Dale's Law has been studied in classical SNN literature \citep{Cornford2021,Cortes2013,Balwani2025} but rarely combined with parallelizable training. Tsodyks--Markram STP \citep{TsodyksMarkram1997,Mongillo2008} has resisted parallel integration due to facilitation--recovery coupling. STDP \citep{FremauxGerstner2016} has typically been integrated only within sequential SNN simulators. To date no framework combines all of these biological priors within an $\bigO(\log L)$ parallel-trainable SSM.

\section{Methodology}
\label{sec:method}

\begin{figure}[H]
\centering
\includegraphics[width=\linewidth]{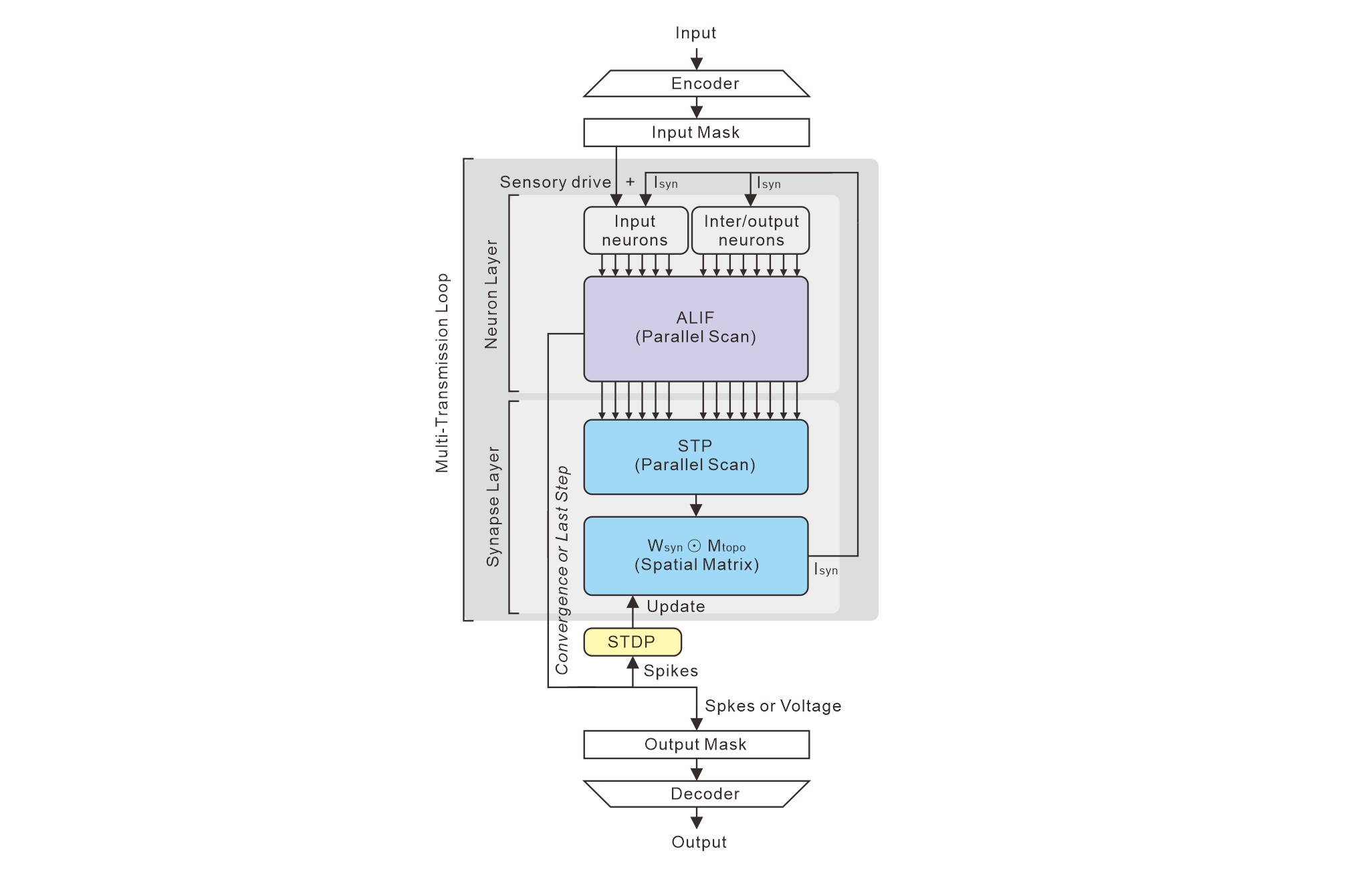}
\caption{Detailed signal flow of the PHCSSM forward pass. The input sequence is projected via a linear encoder and gated by an input mask restricting sensory drive to designated populations. Within the MTL, the NL performs three sequential diagonal parallel scans (membrane potential, adaptive threshold, and refractory suppression) followed by pointwise spike generation (ALIF). The SL applies a synaptic delay buffer, modulates spikes via Tsodyks--Markram STP (two additional parallel scans), and transmits the result through the biologically constrained weight matrix $\Wstruct = \Wsyn \odot \Mtopo$. Convergence is assessed via the Cauchy criterion after each transmission; upon exit, STDP updates synaptic weights using binary spike timing. The output membrane voltage is gated by an output mask and decoded via a linear readout.}
\label{fig:fig3}
\end{figure}

PHCSSM transforms an input sequence $X \in \R^{B \times T \times D_{\mathrm{in}}}$ into class logits $y \in \R^{B \times C}$, with $B$ the batch size, $T$ the sequence length, $D_{\mathrm{in}}$ the input dimensionality, $D$ the neuron dimension, and $C$ the number of output classes. The forward pass proceeds in three operational stages (Figure~\ref{fig:fig3}): a sensory encoder projects $X$ into the neuron dimension; a within-timestep MTL iterates $\Nmax$ times between a Neuron Layer (NL) operator that integrates per-neuron temporal dynamics and a Synapse Layer (SL) operator that mediates inter-neuronal communication; and a readout aggregates the final-iteration activity into $y$. The SL is functionally decomposed into a Pre-synapse Module that handles per-neuron pre-synaptic processing (synaptic delay and short-term plasticity dynamics, both expressed as diagonal parallel scans) and a Post-synapse Module that effects the spatial transmission step through a single matrix multiplication by the topology-masked, Dale-clamped recurrent weight $\Wstruct$. Every temporal recurrence inside NL and the Pre-synapse Module admits a closed-form affine reformulation and is evaluated in $\bigO(\log T)$ parallel work-depth via a log-domain prefix scan, leaving the $\Nmax$-bounded MTL iteration as the only sequential bottleneck per training step.

Training combines gradient-based optimisation of a composite supervised-plus-regulariser loss with a complementary STDP update. Both pathways modify the recurrent weight $\Wsyn$, and each composes with a post-update Dale sign clamp before the next forward pass.

\subsection{Sensory Encoder and Initial Carry}
\label{sec:method-encoder}

The raw input sequence $X \in \R^{B \times T \times D_{\mathrm{in}}}$ is projected into the neuron dimension via a linear encoder with layer normalisation, then gated by a binary input mask $\Min \in \{0,1\}^{D}$ that restricts sensory drive to designated input populations:
\begin{equation}
x_{\mathrm{sen}} \;=\; \rmsnorm\!\left( W_{\mathrm{enc}}\, X + b_{\mathrm{enc}} \right) \odot \Min,
\label{eq:enc}
\end{equation}
where $W_{\mathrm{enc}} \in \R^{D \times D_{\mathrm{in}}}$ and $b_{\mathrm{enc}} \in \R^{D}$ are learnable. The resulting $x_{\mathrm{sen}} \in \R^{B \times T \times D}$ is the persistent sensory drive that re-enters the loop at every transmission iteration (Eq.~\ref{eq:mtl-recirc}). The first iteration receives a pre-scaled copy of the sensory drive as its initial loop input:
\begin{equation}
I^{(1)} \;=\; \alpha_{\mathrm{drive}}\, x_{\mathrm{sen}},
\label{eq:init-carry}
\end{equation}
where $\alpha_{\mathrm{drive}} > 0$ is a scalar pre-synaptic scaling that controls the persistent-drive strength. Applying $\alpha_{\mathrm{drive}}$ from the first iteration is consistent with its physiological role as a fixed input-gain factor and yields a symmetric treatment of the sensory drive across all loop iterations.

\subsection{Neuron Layer (NL): Intrinsic Membrane Dynamics}
\label{sec:method-NL}

The NL encapsulates per-neuron temporal dynamics as an ALIF formulation \citep{Bellec2018,Teeter2018}, strictly diagonal across the neuron dimension. The NL maps an input current $I \in \R^{B \times T \times D}$ to an output spike train $s \in \{0,1\}^{B \times T \times D}$ and the final membrane potential $\Vmem \in \R^{B \times T \times D}$. Within the MTL, $I$ equals $I^{(1)}$ from Eq.~(\ref{eq:init-carry}) at $k=1$ and the recirculated $I^{(k)}$ from Eq.~(\ref{eq:mtl-recirc}) for $k>1$.

The first scan integrates the loop input $I$ through a learnable excitatory decay:
\begin{equation}
V^{\mathrm{exc}}_{t} \;=\; \alpha^{\mathrm{exc}}_{t}\, V^{\mathrm{exc}}_{t-1} \;+\; \softplus(I_{t}),
\qquad \alpha^{\mathrm{exc}}_{t} \;=\; 0.99\,\sigma(\tau_{\mathrm{exc}}),
\label{eq:nl-scan1}
\end{equation}
where $\tau_{\mathrm{exc}}$ is a learnable per-neuron time-constant logit, $\sigma$ is the logistic sigmoid, and the factor $0.99$ ensures strict contractivity $|\alpha^{\mathrm{exc}}_{t}| < 1$ by capping the decay below unity. The $\softplus(\cdot)$ transform guarantees positivity of the scan input. The initial state $V^{\mathrm{exc}}_{0}=0$ is restored at the start of every transmission iteration, since the neuron and synapse states are reset at each iteration of the MTL.

The second scan computes an adaptive firing threshold driven by the membrane-to-threshold proximity:
\begin{equation}
\eta_{t} \;=\; \alpha^{\mathrm{adapt}}\, \eta_{t-1} \;+\; \sigma\!\left( V^{\mathrm{exc}}_{t} - \Vthstat \right),
\qquad \alpha^{\mathrm{adapt}} = \sigma(\tau_{\mathrm{adapt}}),
\label{eq:nl-scan2}
\end{equation}
where $\tau_{\mathrm{adapt}}$ is the threshold-adaptation time-constant, $\Vthstat$ is a learnable firing threshold, and the scan input $\sigma(\cdot)$ saturates within $(0,1)$ so that the adaptation contribution is bounded. The initial state is $\eta_{0}=0$.

The static threshold combines with the adaptive contribution to yield the effective threshold:
\begin{equation}
V_{\mathrm{th},t} \;=\; \Vthstat \;+\; \beta\, \eta_{t},
\label{eq:vth}
\end{equation}
where $\beta$ is a learnable scalar that weights the adaptation term. The preliminary spike train is obtained by Heaviside thresholding:
\begin{equation}
s^{\mathrm{pre}}_{t} \;=\; \heaviside\!\left( V^{\mathrm{exc}}_{t} - V_{\mathrm{th},t} \right),
\label{eq:spre}
\end{equation}
where $\heaviside$ is the Heaviside step function. The preliminary spike $s^{\mathrm{pre}}$ feeds the refractory scan below; the final spike emitted by the NL is computed in Eq.~(\ref{eq:nl-out}) after subtracting the refractory contribution.

The third scan models post-spike refractory decay:
\begin{equation}
V_{\mathrm{res},t} \;=\; \alpha^{\mathrm{ref}}_{t}\, V_{\mathrm{res},t-1} \;+\; \softplus\!\left( s^{\mathrm{pre}}_{t-1}\, w_{\mathrm{reset}} \right),
\qquad \alpha^{\mathrm{ref}}_{t} = 0.99\,\sigma(\tau_{\mathrm{ref}}),
\label{eq:nl-scan3}
\end{equation}
where $\tau_{\mathrm{ref}}$ is the refractory time-constant and $w_{\mathrm{reset}}$ is a learnable per-neuron reset weight. As in Scan~1, the factor $0.99$ caps $|\alpha^{\mathrm{ref}}_{t}| < 1$ below unity, and $\softplus(\cdot)$ enforces positivity. The initial state is $V_{\mathrm{res},0}=0$.

The NL emits the refractory-corrected membrane potential and the corresponding spike:
\begin{equation}
V_{\mathrm{mem},t} \;=\; V^{\mathrm{exc}}_{t} - V_{\mathrm{res},t},
\qquad
s_{t} \;=\; \heaviside\!\left( V_{\mathrm{mem},t} - V_{\mathrm{th},t} \right).
\label{eq:nl-out}
\end{equation}
Here $V_{\mathrm{mem},t}$ is the variable used both for downstream readout (when readout-source is voltage) and for the convergence- and voltage-regulariser losses (Section~\ref{sec:method-loss}). The output spike $s_{t}$ is the variable consumed by the synapse-layer delay buffer (Eq.~\ref{eq:delay}), by the STDP eligibility traces, and as the rate-regulariser argument.

\subsection{Synapse Layer (SL): Constrained Inter-Neuronal Connectivity}
\label{sec:method-SL}

All inter-neuron communication is delegated to the SL, invoked once per transmission iteration after the NL. The SL is functionally partitioned into a Pre-synapse Module (synaptic delay and STP dynamics, both expressed as diagonal parallel scans) and a Post-synapse Module that effects the spatial lateral communication step via a single matrix multiplication through the topology-masked, Dale-clamped recurrent weight $\Wstruct$. The SL maps $s_{t}$ (Eq.~\ref{eq:nl-out}) to the synaptic current $I_{\mathrm{syn},t}$ that re-enters the MTL via Eq.~(\ref{eq:mtl-recirc}); its state is reset at the start of every iteration.

Pre-synaptic spikes pass through a First-In-First-Out (FIFO) ring buffer of depth $d$ before reaching the STP and post-synapse stages:
\begin{equation}
s^{d}_{t} \;=\; s_{t-d},
\label{eq:delay}
\end{equation}
where $d \in \mathbb{Z}_{\ge 0}$ is a fixed integer delay (default $d=1$) and $s^{d}_{t}$ denotes the delayed spike train. The delay introduces a minimal model of axonal propagation with no learnable parameters; it is omitted ($d=0$) when the dataset does not benefit from explicit delay.

A Tsodyks--Markram STP model \citep{TsodyksMarkram1997} is implemented via two per-neuron state variables: a facilitation variable $u_{t}$ and a recovery variable $x_{t}$. A generalised affine form of the discrete update equations is adopted, exactly absorbing both the spike-conditioned facilitation jump and the spike-conditioned recovery depletion into a single time-varying affine recurrence per variable. This formulation makes both variables parallelisable in $\bigO(\log T)$ using a two-coefficient extension of the scan primitive.

The facilitation variable obeys:
\begin{equation}
u_{t} \;=\; \big(1 - \alpha^{u}_{t} \cdot U_{\mathrm{amp}}\, s^{d}_{t}\big)\, \alpha^{u}_{t}\, u_{t-1} \;+\; \big(1 - \alpha^{u}_{t}\big)\, U_{0} \;+\; \alpha^{u}_{t}\, U_{\mathrm{amp}}\, s^{d}_{t},
\label{eq:stp-u}
\end{equation}
where $\alpha^{u}_{t} = \exp(-\Delta t / \tau_{f})$, $\tau_{f}$ is the per-neuron facilitation time-constant, $U_{0}$ is a learnable per-neuron baseline release probability, and $U_{\mathrm{amp}}$ is a fixed global facilitation amplitude. The recovery variable obeys:
\begin{equation}
x_{t} \;=\; \big(1 - u_{t}\, s^{d}_{t}\big)\, \alpha^{x}_{t}\, x_{t-1} \;+\; \big(1 - \alpha^{x}_{t}\big),
\label{eq:stp-x}
\end{equation}
where $\alpha^{x}_{t} = \exp(-\Delta t / \tau_{d})$ with $\tau_{d}$ the recovery time-constant. The $\alpha$-coefficient $(1-\alpha^{u}_{t}\, U_{\mathrm{amp}}\, s^{d}_{t})\,\alpha^{u}_{t}$ in Eq.~(\ref{eq:stp-u}) and $(1-u_{t}\, s^{d}_{t})\,\alpha^{x}_{t}$ in Eq.~(\ref{eq:stp-x}) are clamped to $[0,1)$ inline before the scan to remain inside the valid contractive range; after the scans, both variables are clipped to the biological range $[0,1]$:
\begin{equation}
u_{t} \;\leftarrow\; \mathrm{clip}(u_{t},\,0,\,1),\qquad x_{t} \;\leftarrow\; \mathrm{clip}(x_{t},\,0,\,1).
\label{eq:stp-clip}
\end{equation}
The clamping prevents numerical excursion outside $[0,1]$ under boundary cases (e.g., when the facilitation-amplitude factor $U_{\mathrm{amp}}$ temporarily drives the facilitation increment near unity).

The two STP variables combine into a per-time-step, per-neuron multiplicative gating factor:
\begin{equation}
g^{\mathrm{stp}}_{t} \;=\; u_{t}\, x_{t},
\label{eq:stp-gate}
\end{equation}
which acts as a dynamic, time-varying scaling of the pre-synaptic spike train before the post-synaptic projection (Eq.~\ref{eq:isyn}). By construction, $g^{\mathrm{stp}}_{t} \in [0,1]$.

The synaptic current produced by the SL at the end of one transmission iteration is:
\begin{equation}
I_{\mathrm{syn},t} \;=\; \Wstruct \cdot \big( g^{\mathrm{stp}}_{t} \odot s^{d}_{t} \big).
\label{eq:isyn}
\end{equation}
Scaling the pre-synaptic spike train by $g^{\mathrm{stp}}_{t}$ prior to the matrix multiplication implements a time-varying effective weight matrix $\Wstruct \cdot \mathrm{diag}(g^{\mathrm{stp}}_{t})$, transforming the static structural connectivity $\Wstruct$ (defined next) into state-dependent effective weights without an explicit dynamic-weight tensor.

\subsection{Recurrent Weight $\Wsyn$: Lifecycle and Biological Priors}
\label{sec:method-Wsyn}

The recurrent synaptic weight $\Wsyn \in \R^{D \times D}$ is the only trainable parameter carrying cross-neuron coupling. Two biological priors act on it: a fixed topology mask $\Mtopo \in \{0,1\}^{D\times D}$ applied as a forward-pass multiplicative mask, and Dale's sign clamp applied as a post-update parameter projection. They compose into the effective recurrent weight $\Wstruct$ consumed by Eq.~(\ref{eq:isyn}) at every forward step.

At model construction, $\Wsyn$ is initialised with fan-in-scaled Gaussian noise $\mathcal{N}(0,\,1/D)$. The topology mask $\Mtopo$ is constructed once at the same time from the hierarchical-connectome specification (number of macro-regions $R$, per-region E/I ratios, and pairwise inter-region connection probabilities). $\Mtopo$ is not trained; it remains fixed across the entire training run.

Dale's law \citep{StrataHarvey1999} requires that all outgoing synapses from a single neuron share the same neurotransmitter sign: excitatory neurons project with non-negative weights, inhibitory neurons with non-positive weights. PHCSSM enforces this as a post-update parameter projection applied immediately after every parameter update on $\Wsyn$:
\begin{equation}
\Wsyn[:,E] \;\leftarrow\; \max(\Wsyn[:,E],\,0),
\qquad
\Wsyn[:,I] \;\leftarrow\; \min(\Wsyn[:,I],\,0),
\label{eq:dale}
\end{equation}
where $\Wsyn[:,E]$ denotes the columns indexed by excitatory neurons (clipped to $\ge 0$ element-wise) and $\Wsyn[:,I]$ denotes the columns indexed by inhibitory neurons (clipped to $\le 0$ element-wise). The clamp is applied twice per minibatch: once after the gradient step on $\Wsyn$ (Eq.~\ref{eq:adam}) and once after the STDP additive update (Eq.~\ref{eq:rstdp}). It acts as an in-place projection of $\Wsyn$ onto the Dale-consistent subspace, executed outside the autograd graph.

At every transmission iteration of the forward pass, the Dale-clamped $\Wsyn$ is multiplied element-wise by $\Mtopo$ to yield the effective recurrent weight matrix:
\begin{equation}
\Wstruct \;=\; \Wsyn \odot \Mtopo,
\label{eq:wstruct}
\end{equation}
which is the matrix consumed by Eq.~(\ref{eq:isyn}) inside the SL. Because $\Mtopo$ is fixed and Dale's clamp has already been applied to $\Wsyn$ after the previous parameter update, no clamping is required at forward time; $\Wstruct$ is simply read off. Different choices of $\Mtopo$ (e.g., a two-region hierarchical specification versus a single-region random specification) constitute architectural priors and are selected per dataset.

\subsection{Multi-Transmission Loop (MTL) and Spatial Recurrence}
\label{sec:method-MTL}

The diagonal NL keeps neurons within the same timestep mutually decoupled during the parallel temporal scans. To recover spatial coupling without sacrificing time-parallelism, the NL--SL pair is invoked $\Nmax$ times per physical timestep. For $k=1,\dots,\Nmax$, one iteration consists of an NL invocation producing $(\Vmem^{(k)},\,s^{(k)})$ from the loop input $I^{(k)}$ (Eqs.~\ref{eq:nl-scan1}--\ref{eq:nl-out}), followed by an SL invocation producing $\Isyn^{(k)}$ from $s^{(k)}$ (Eqs.~\ref{eq:delay}--\ref{eq:isyn}); the loop input for the next iteration is then updated via Eq.~(\ref{eq:mtl-recirc}). The NL and SL states are reset at the start of every iteration, so that iteration $k=\Nmax$ closes with a single-pass RSNN-equivalent forward.

The relative inter-iteration residual of the synaptic current is monitored during the forward pass:
\begin{equation}
r^{(k)} \;=\; \frac{\|\Isyn^{(k)} - \Isyn^{(k-1)}\|_{2}}{\|\Isyn^{(k-1)}\|_{2} + \epsilon} \;<\; \theta_{\mathrm{conv}},
\label{eq:cauchy}
\end{equation}
where $\epsilon = 10^{-8}$ is a numerical floor and $\theta_{\mathrm{conv}} = 10^{-3}$ is the convergence threshold. The residual $r^{(k)}$ provides an algorithmic stopping criterion $\Mstep = \min\{k : r^{(k)} < \theta_{\mathrm{conv}}\}$. In the present formulation, the forward pass runs for a fixed $\Nmax$ iterations and the convergence-loss regulariser $\Lconv$ penalises iteration-over-iteration spike-output drift in lieu of runtime early-exit. This formulation expresses the entire MTL as a single static computation graph while preserving the convergence-monitoring semantics of the Cauchy criterion.

The MTL input for the next iteration combines the current synaptic current with the re-injected sensory drive:
\begin{equation}
I^{(k+1)} \;=\; \Isyn^{(k)} \;+\; \alpha_{\mathrm{drive}}\, x_{\mathrm{sen}},
\label{eq:mtl-recirc}
\end{equation}
where $x_{\mathrm{sen}}$ is constant across $k$ and acts as a persistent boundary condition. The initial loop input at $k=1$ is $I^{(1)} = \alpha_{\mathrm{drive}}\, x_{\mathrm{sen}}$ (Eq.~\ref{eq:init-carry}).

\subsection{Readout}
\label{sec:method-readout}

The classifier head consumes the final-iteration activity of the MTL through a configurable aggregation operator and a linear projection to the class space:
\begin{equation}
y \;=\; W_{\mathrm{dec}}\,\mathrm{readout}_{\mathrm{op}}(z^{(\Nmax)}) + b_{\mathrm{dec}},
\label{eq:readout}
\end{equation}
where $\mathrm{readout}_{\mathrm{op}}$ selects the temporal-aggregation strategy from
\{\textit{mean}, \textit{last}, \textit{sum}, \textit{max}, \textit{weighted}, \textit{com}, \textit{ssm}\}:
\textit{mean} (default) averages across time,
\textit{last} returns the final timestep,
\textit{sum} and \textit{max} apply the corresponding reduction,
\textit{weighted} applies a learnable time-weighting,
\textit{com} computes a centre-of-mass index,
and \textit{ssm} applies RMSNorm followed by the final-timestep value.
The argument $z^{(\Nmax)}$ is either the voltage trace $\Vmem^{(\Nmax)}$ or the spike trace $s^{(\Nmax)}$, selected by the readout-source hyper-parameter and chosen per dataset. The parameters $W_{\mathrm{dec}} \in \R^{C \times D}$ and $b_{\mathrm{dec}} \in \R^{C}$ are learnable.

\subsection{Training Objectives}
\label{sec:method-loss}

Training minimises a composite objective combining cross-entropy on the readout logits with three regularisers (spike-rate target deviation, voltage $\ell_{2}$ penalty, and convergence regulariser on iteration-over-iteration spike-output drift):
\begin{equation}
\Ltotal \;=\; \Ltask \;+\; \lambda_{\mathrm{rate}}\,\Lrate \;+\; \lambda_{\mathrm{volt}}\,\Lvolt \;+\; \lambda_{\mathrm{conv}}\,\Lconv,
\label{eq:Ltotal}
\end{equation}
where $\Ltask$ is label-smoothed cross-entropy on the readout logits, and the three regularisers control firing-rate balance, membrane-voltage boundedness, and convergence to the within-step fixed point.

All trainable parameters $\Theta$ are updated by AdamW \citep{LoshchilovHutter2019}. For the recurrent weight $\Wsyn$, the update composes with Dale's clamp (Eq.~\ref{eq:dale}):
\begin{equation}
\Wsyn \;\leftarrow\; \mathrm{Dale}\!\left( \Wsyn - \eta\, \nabla_{\Wsyn}\Ltotal \right).
\label{eq:adam}
\end{equation}

\subsection{Spike-Timing-Dependent Plasticity}
\label{sec:method-rstdp}

An STDP rule provides a second online synaptic learning pathway that runs in parallel with the gradient-based update. The STDP update operates on the same $\Wsyn$, uses the same forward-pass spikes, and is computed outside the autograd graph; it is gated by both $\Mtopo$ and a learnable Hebbian-type mask $\Mhebb$, then composes with Dale's clamp (Eq.~\ref{eq:dale}):
\begin{equation}
\Wsyn \;\leftarrow\; \mathrm{Dale}\!\left( \Wsyn \;+\; \eta_{\mathrm{hebb}}\, \Mtopo \odot \Mhebb \odot \Delta W_{\mathrm{hebb}} \right),
\label{eq:rstdp}
\end{equation}
where $\Delta W_{\mathrm{hebb}}$ aggregates an asymmetric pre-post Hebbian increment over the sequence via exponentially-decaying LTP and LTD eligibility traces.

\subsection{Asymptotic Equivalence to Sequential RSNN}
\label{sec:method-equiv}

PHCSSM supports two execution modes that share the same trained weights and per-step operators (Eqs.~\ref{eq:nl-scan1}--\ref{eq:isyn}) but schedule them differently.

\emph{Training mode} (Algorithm~\ref{alg:phcssm}) makes the within-step iteration $k = 1,\dots,\Nmax$ the outer loop. Within each $k$, NL and SL are each applied once to the entire $T$-length sequence via the log-depth parallel scan of Eq.~(\ref{eq:nl-scan1}); their internal state is reset to zero at the start of every $k$, with only the loop input $I^{(k)}$ (Eq.~\ref{eq:mtl-recirc}) propagated between iterations. The $\Nmax$-th iteration's output feeds the readout. \emph{RSNN deployment mode} instead makes the timestep $t = 0,\dots,T-1$ the outer loop and applies NL$+$SL exactly once per timestep, with the full per-neuron and per-synapse state carried over from $t-1$. There is no within-step iteration loop and no log-domain scan.

This loop-axis swap distinguishes PHCSSM from the standard SSM training-versus-inference distinction, where both modes loop over $t$ and differ only in parallel-scan versus sequential execution. The $\Nmax$ within-step iterations of PHCSSM's training mode are necessary to refine the lateral coupling that the diagonal parallel scan over $T$ would otherwise leave unresolved; at deployment, this within-step refinement is absorbed into per-timestep state carry-over (Figure~\ref{fig:fig1}B), eliminating the $K$ loop entirely while keeping the trained weights (Figure~\ref{fig:fig1}C) and biological constraints intact.

Under the contractive regime enforced by the convergence regulariser $\Lconv$, the training-mode iteration converges as $k \to \Nmax$ to the fixed point of the per-step recurrence. The per-iteration state reset is what aligns the two modes: it ensures the $\Nmax$-th iteration applies the same operator stack on the same zero-state initial conditions a chip-deployed RSNN single pass also sees, while the persistent sensory drive (Eq.~\ref{eq:mtl-recirc}) prevents the fixed point from diluting as $\Nmax$ grows. In exact arithmetic the two trajectories coincide; in finite IEEE-754 precision they differ only by floating-point reduction-order rounding.

The parallel-scan training mode pays an $\Nmax$-fold compute overhead per timestep but is required for tractable BPTT on long sequences. The sequential RSNN deployment mode removes the within-step iteration loop entirely, yielding a per-inference compute reduction proportional to $\Nmax$. Because the trained weights are biologically constrained throughout, the RSNN deployment form is a biologically constrained RSNN by construction, without the conversion error of standard ANN-to-SNN pipelines.

\subsection{Implementation}
\label{sec:method-impl}

PHCSSM is implemented in Python using JAX \citep{Bradbury2018} with Flax and Equinox \citep{KidgerGarcia2021} for parameter management and the AdamW optimiser provided by optax \citep{DeepMindJAX2020}. The log-domain prefix scan and the MTL are expressed as a single compiled static computation graph, with all forward and backward operations of one training step executing as one accelerated kernel sequence in float32. Five-seed training and accuracy evaluation use a single NVIDIA H100 GPU; training-time benchmarking uses an NVIDIA RTX 4090 to match the baseline timing protocol; cross-backend reproducibility additionally uses a single-thread x86 CPU, an ARM Cortex-A76 (Raspberry Pi 5), and the STM32L412KB Cortex-M4F microcontroller (40~KB SRAM, 128~KB Flash).

\begin{algorithm}[t]
\caption{PHCSSM Forward Pass}
\label{alg:phcssm}
\begin{algorithmic}[1]
\Require Input sequence $X \in \R^{B \times T \times D_{\mathrm{in}}}$, parameters $\Theta$
\Ensure Membrane voltage $v \in \R^{B \times T \times D}$, spike train $s \in \{0,1\}^{B \times T \times D}$
\State $x_{\mathrm{sen}} \gets \mathrm{Encode}(X)$ \Comment{Linear + LayerNorm + input mask}
\State $I \gets \alpha_{\mathrm{drive}} \cdot x_{\mathrm{sen}}$
\For{$k = 1$ to $\Nmax$}\Comment{Multi-Transmission Loop}
    \State Reset NL and SL state \Comment{State reset per iteration}
    \State $v,\,s \gets \mathrm{ALIF}(I)$ \Comment{Parallel scan over $T$ timesteps}
    \State $s^{d} \gets \mathrm{Delay}(s,\,d)$ \Comment{Synaptic delay buffer}
    \State $s^{\mathrm{eff}} \gets \mathrm{STP}(s^{d})$ \Comment{Tsodyks--Markram modulated spikes}
    \State $\Isyn \gets \Wstruct \cdot s^{\mathrm{eff}}$ \Comment{Bio-constrained propagation}
    \State $I \gets \Isyn + \alpha_{\mathrm{drive}}\, x_{\mathrm{sen}}$ \Comment{Recirculate with sensory drive}
\EndFor
\If{training}
    \State $\Wstruct \gets \mathrm{STDP}(\Wstruct,\,s)$ \Comment{Hebbian update}
\EndIf
\State \Return $v,\,s$
\end{algorithmic}
\end{algorithm}

\section{Experiments}
\label{sec:exp}

\subsection{Experimental Setup}
\label{sec:exp-setup}

PHCSSM is evaluated on six UEA-MTSCA physiological benchmarks following the \citet{Walker2024} protocol adopted by \citet{RuschRus2025} and \citet{Farsang2025}: Heartbeat ($T=405$, 61~channels, 2~classes), SelfRegulationSCP1 ($T=896$, 6~channels, 2~classes), SelfRegulationSCP2 ($T=1{,}152$, 7~channels, 2~classes), EthanolConcentration ($T=1{,}751$, 3~channels, 4~classes), MotorImagery ($T=3{,}000$, 64~channels, 2~classes), and EigenWorms ($T=17{,}984$, 6~channels, 5~classes). A 70/15/15 train/validation/test split is used with the five fixed seeds $\{2345, 3456, 4567, 5678, 6789\}$ shared with \citet{Walker2024}, \citet{RuschRus2025}, and \citet{Farsang2025}. Configuration selection uses a single-seed grid search over $\{\text{learning rate} \in \{10^{-4}, 10^{-3}\},\,\text{neuron dimension} \in \{16, 32, 64\},\,\text{readout} \in \{\text{voltage}, \text{spike}\},\,\text{topology} \in \{\text{feedforward}, \text{bidirectional}\}\}$, selecting the configuration with the highest validation accuracy; ties at the grid-search seed are resolved by validation accuracy under a second seed. The selected configuration is then re-trained under the full five-seed reporting set, and the headline test accuracy is reported as mean~$\pm$~std over the five seeds. All experiments use a 2-region architecture with E/I ratio 0.8 and fixed STDP parameters ($\tau_{+} = 10$, $\tau_{-} = 20$, $A_{+} = 1.0$, $A_{-} = 1.05$, $\eta_{\mathrm{hebb}} = 0.01$). All measurements use a JAX/Flax implementation on a H100 GPU; cross-backend reproducibility is verified on x86 CPU and Cortex-A76.

Comparisons are made against NRDE, NCDE, Log-NCDE, LRU, S5, S6, Mamba, LinOSS-IMEX, LinOSS-IM, Transformer, and RFormer (results from \citet{RuschRus2025}); LrcSSM (results from \citet{Farsang2025}); and PD-SSM (results from \citet{Terzic2025}). All baselines are unconstrained models without Dale's Law, STP, or STDP; PHCSSM is the only entry satisfying all five biological constraints simultaneously.

\subsection{Classification Accuracy}
\label{sec:exp-acc}

On six UEA physiological benchmarks (Table~\ref{tab:tab1}), PHCSSM achieves competitive test accuracy across the full difficulty spectrum while satisfying five simultaneous neuro-physical constraints. On SelfRegulationSCP2, PHCSSM reaches $58.3 \pm 6.3$\% under $M_{\mathrm{topo}}$~1, statistically tied with the leading SSM baseline LinOSS-IMEX ($58.9 \pm 8.1$\%); the heavily overlapping confidence intervals place PHCSSM among the leaders on this benchmark. On MotorImagery, PHCSSM at $54.7 \pm 4.5$\% surpasses Mamba, Transformer, NCDE, LRU, S6, and S5 (which span 47.7 to 53.0\%). On EigenWorms ($T=17{,}984$), PHCSSM at $85.0 \pm 5.8$\% surpasses LinOSS-IMEX, Mamba, Log-NCDE, NCDE, NRDE, and S5, and matches LRU and S6. On Heartbeat, PHCSSM at $73.9 \pm 3.9$\% matches NRDE and exceeds LrcSSM, RFormer, Transformer, and NCDE, providing a stable result at the shortest-sequence extreme of the benchmark. On SelfRegulationSCP1 ($80.7 \pm 1.6$\%) and EthanolConcentration ($31.1 \pm 6.1$\%), PHCSSM remains competitive with the SSM family. PHCSSM is the only model in the comparison achieving this competitive standing across all six datasets while integrating ALIF dynamics, synaptic delay, Tsodyks--Markram short-term plasticity, Dale's Law with E/I-asymmetric topology, and STDP online learning.

\begin{table}[!htbp]
\centering
\caption{Test accuracy (mean~$\pm$~std over 5 seeds) on UEA physiological benchmarks. $\dagger$~Results from \citet{RuschRus2025}. $\ddagger$~Results from \citet{Farsang2025}. *Results from \citet{Terzic2025}. Best per column in bold. PHCSSM is reported under two topology matrix configurations ($M_{1}$ and $M_{2}$); the best result per dataset across both configurations is underlined and used in all per-dataset comparisons. Gray background indicates accuracy lower than PHCSSM.}
\label{tab:tab1}
\scriptsize
\setlength{\tabcolsep}{4pt}
\begin{tabular}{l c c c c c c}
\toprule
Model & Heartbeat & SCP1 & SCP2 & EthanolConc. & Motor-Im. & EigenWorms \\
Seq. length & 405 & 896 & 1{,}152 & 1{,}751 & 3{,}000 & 17{,}984 \\
Channels & 61 & 6 & 7 & 2 & 64 & 6 \\
Classes & 2 & 2 & 2 & 4 & 2 & 5 \\
\midrule
NRDE$^{\dagger}$              & 73.9$\pm$2.6 & \cellcolor{tabgray}76.7$\pm$5.6 & \cellcolor{tabgray}48.1$\pm$11.4 & 31.4$\pm$4.5 & \cellcolor{tabgray}54.0$\pm$7.8 & \cellcolor{tabgray}77.2$\pm$7.1 \\
NCDE$^{\dagger}$              & \cellcolor{tabgray}68.1$\pm$5.8 & \cellcolor{tabgray}80.0$\pm$2.0 & \cellcolor{tabgray}49.1$\pm$6.2 & \cellcolor{tabgray}22.0$\pm$1.0 & \cellcolor{tabgray}51.6$\pm$6.2 & \cellcolor{tabgray}62.2$\pm$2.2 \\
Log-NCDE$^{\dagger}$          & 74.2$\pm$2.0 & 82.1$\pm$1.4 & \cellcolor{tabgray}54.0$\pm$2.6 & 35.9$\pm$6.1 & 57.2$\pm$5.6 & \cellcolor{tabgray}82.8$\pm$2.7 \\
LRU$^{\dagger}$               & 78.1$\pm$7.6 & 84.5$\pm$4.6 & \cellcolor{tabgray}47.4$\pm$4.0 & \cellcolor{tabgray}23.8$\pm$2.8 & \cellcolor{tabgray}51.9$\pm$8.6 & \cellcolor{tabgray}85.0$\pm$6.2 \\
S5$^{\dagger}$                & 73.9$\pm$3.1 & 87.1$\pm$2.1 & \cellcolor{tabgray}55.1$\pm$3.3 & \cellcolor{tabgray}25.6$\pm$3.5 & \cellcolor{tabgray}53.0$\pm$3.9 & \cellcolor{tabgray}83.9$\pm$4.1 \\
S6$^{\dagger}$                & 76.5$\pm$8.3 & 82.8$\pm$2.7 & \cellcolor{tabgray}49.9$\pm$9.4 & \cellcolor{tabgray}26.4$\pm$6.4 & \cellcolor{tabgray}51.3$\pm$4.7 & \cellcolor{tabgray}85.0$\pm$16.1 \\
Mamba$^{\dagger}$             & 76.2$\pm$3.8 & 80.7$\pm$1.4 & \cellcolor{tabgray}48.2$\pm$3.9 & \cellcolor{tabgray}27.9$\pm$4.5 & \cellcolor{tabgray}47.7$\pm$4.5 & \cellcolor{tabgray}70.9$\pm$15.8 \\
LinOSS-IMEX$^{\dagger}$       & 75.5$\pm$4.3 & 87.5$\pm$4.0 & 58.9$\pm$8.1 & \cellcolor{tabgray}29.9$\pm$1.0 & 57.9$\pm$5.3 & \cellcolor{tabgray}80.0$\pm$2.7 \\
LinOSS-IM$^{\dagger}$         & 75.8$\pm$3.7 & 87.8$\pm$2.6 & \cellcolor{tabgray}58.2$\pm$6.9 & \cellcolor{tabgray}29.9$\pm$0.6 & 60.0$\pm$7.5 & 95.0$\pm$4.4 \\
Transformer$^{\ddagger}$      & \cellcolor{tabgray}70.5$\pm$0.1 & 84.3$\pm$6.3 & \cellcolor{tabgray}49.1$\pm$2.5 & 40.5$\pm$6.3 & \cellcolor{tabgray}50.5$\pm$3.0 & \cellcolor{tabgray}OOM \\
RFormer$^{\ddagger}$          & \cellcolor{tabgray}72.5$\pm$0.1 & 81.2$\pm$2.8 & \cellcolor{tabgray}52.3$\pm$3.7 & 34.7$\pm$4.1 & 55.8$\pm$6.6 & 90.3$\pm$0.1 \\
LrcSSM$^{\ddagger}$           & \cellcolor{tabgray}72.7$\pm$5.7 & 85.2$\pm$2.1 & \cellcolor{tabgray}53.9$\pm$7.2 & 36.9$\pm$5.3 & 58.6$\pm$3.1 & 90.6$\pm$1.4 \\
PD-SSM$^{*}$                  & 80.0$\pm$2.6 & 80.9$\pm$2.0 & \cellcolor{tabgray}56.1$\pm$8.6 & 34.7$\pm$4.0 & 60.0$\pm$3.7 & 90.0$\pm$5.7 \\
\midrule
PHCSSM ($M_{topo}1$) & \underline{73.9$\pm$3.9} & \underline{80.7$\pm$1.6} & 53.3$\pm$6.8 & 30.4$\pm$6.7 & 53.3$\pm$4.0 & 84.4$\pm$5.8 \\
PHCSSM ($M_{topo}2$) & 71.3$\pm$6.5 & 80.5$\pm$2.8 & \underline{58.3$\pm$6.3} & \underline{31.1$\pm$6.1} & \underline{54.7$\pm$4.5} & \underline{85.0$\pm$5.8} \\
\bottomrule
\end{tabular}
\end{table}

\subsection{Parameter and Training Cost}
\label{sec:exp-cost}

PHCSSM's parameter cost is $\BigTheta(D^{2})$, one factor of $L$ below the $\BigTheta(D^{2}L)$ of $L$-layer stacked diagonal SSMs (Table~\ref{tab:tab2}), because the PHC framework shares the connectome and diagonal-core parameters across all hierarchical regions rather than instantiating an independent block per layer. The empirical consequence (Table~\ref{tab:tab3}) is that PHCSSM is the smallest model in the comparison on every dataset by one to three orders of magnitude. Its trainable parameter count ranges from 1{,}312 on Heartbeat to 4{,}891 on EigenWorms; the smallest unconstrained baseline, S6, requires 4 to 10 times more parameters per dataset (5{,}780 to 52{,}802), and the largest model in the comparison, NRDE on Heartbeat, reaches 15.7 million, four orders of magnitude above PHCSSM. Even relative to the parameter-efficient LinOSS family, PHCSSM uses 4 to 50 times fewer trainable parameters per dataset.

\begin{table}[!htbp]
\centering
\caption{Architectural complexity comparison of SSM variants. PHCSSM's NL is analogous to the diagonal SSM core in LRU, S5, Mamba, LinOSS, and LrcSSM, while the SL is analogous to the inter-layer MLP. Because NL and SL are shared across all $M$ transmission steps, total parameter count is $\BigTheta(D^{2})$ rather than $\BigTheta(D^{2}L)$ for $L$-layer stacked architectures.}
\label{tab:tab2}
\scriptsize
\setlength{\tabcolsep}{5pt}
\begin{tabular}{l l l l l l}
\toprule
Model    & Recurrence              & Params         & Train time      & Memory & Bio. constraints \\
\midrule
LRU      & Temporal (linear)       & $\BigTheta(D^{2}L)$ & $\bigO(T D^{2} L)$ & $\bigO(T D)$ & None \\
S5       & Temporal (linear)       & $\BigTheta(D^{2}L)$ & $\bigO(T D^{2} L)$ & $\bigO(T D)$ & None \\
Mamba    & Temporal (selective)    & $\BigTheta(D^{2}L)$ & $\bigO(T D^{2} L)$ & $\bigO(T D)$ & None \\
LinOSS   & Temporal (oscillatory)  & $\BigTheta(D^{2}L)$ & $\bigO(T D^{2} L)$ & $\bigO(T D)$ & None \\
LrcSSM   & Temporal (nonlinear)    & $\BigTheta(D^{2}L)$ & $\bigO(T D^{2} L)$ & $\bigO(T D)$ & Partial (LTC) \\
PHCSSM   & Spatiotemporal          & $\BigTheta(D^{2})$  & $\bigO(T D^{2} M)$ & $\bigO(T D)$ & Full (5 constraints) \\
\bottomrule
\end{tabular}
\end{table}

\begin{table}[!htbp]
\centering
\caption{Number of parameters for every considered model on all long-range datasets. GPU memory reflects peak activation during backpropagation. All measurements were performed on an NVIDIA RTX~4090 GPU using JAX. Values for all baseline models are from $\dagger$~\citet{RuschRus2025}.}
\label{tab:tab3}
\scriptsize  
\setlength{\tabcolsep}{4pt}
\begin{tabular}{l r r r r r r}
\toprule
Model                     & Heartbeat   & SCP1     & SCP2    & Ethanol   & Motor    & Worms   \\
\midrule
NRDE$^{\dagger}$          & 15{,}657{,}742 & 117{,}187  & 200{,}707 & 93{,}212    & 1{,}134{,}395 & 105{,}110 \\
NCDE$^{\dagger}$          & 1{,}098{,}114  & 166{,}274  & 182{,}914 & 133{,}252   & 186{,}962    & 166{,}789 \\
Log-NCDE$^{\dagger}$      & 168{,}320    & 91{,}557  & 36{,}379  & 31{,}452    & 81{,}391     & 37{,}977  \\
LRU$^{\dagger}$           & 338{,}820    & 25{,}892  & 26{,}020  & 76{,}522    & 107{,}544    & 101{,}129 \\
S5$^{\dagger}$            & 158{,}310    & 226{,}328 & 5{,}652   & 76{,}214    & 17{,}496     & 22{,}007  \\
Mamba$^{\dagger}$         & 1{,}034{,}242  & 184{,}194  & 356{,}290 & 1{,}032{,}772 & 228{,}226    & 27{,}381  \\
S6$^{\dagger}$            & 6{,}674     & 24{,}898  & 26{,}018  & 5{,}780     & 52{,}802     & 15{,}045  \\
LinOSS-IMEX$^{\dagger}$   & 29{,}444    & 447{,}944 & 448{,}072 & 70{,}088    & 106{,}024    & 26{,}119  \\
LinOSS-IM$^{\dagger}$     & 10{,}936    & 991{,}240 & 399{,}112 & 6{,}728     & 91{,}844     & 134{,}279 \\
\midrule
\textbf{PHCSSM}           & \textbf{1{,}312} & \textbf{3{,}388} & \textbf{4{,}877} & \textbf{4{,}673} & \textbf{1{,}444} & \textbf{4{,}891} \\
\bottomrule
\end{tabular}
\end{table}

Training wall-clock measurements (Table~\ref{tab:tab4}) for PHCSSM are 18 to 141 seconds per 1000 training steps with STDP enabled, and 15 to 131 seconds with STDP disabled (a 15 to 23 percent reduction in the wall-clock cost). This places PHCSSM within the sub-150-second-per-1000-steps range, comparable to modern parallel-scan diagonal-SSM baselines (combined range 3 to 128 s) and one to two orders of magnitude faster than older ODE-based models (NRDE and Log-NCDE; combined range 583 to 9{,}539 s). PHCSSM was benchmarked on a single NVIDIA RTX 4090 GPU to match the hardware used by \citet{RuschRus2025} for the baseline run-time values cited in Table~\ref{tab:tab4}, ensuring the comparison is on equal compute footing rather than being inflated by a server-class GPU advantage. 

\begin{table}[!htbp]
\centering
\caption{Run time in seconds for the considered models for 1000 training steps. All measurements were performed on an NVIDIA RTX~4090 GPU using JAX. Values for all baseline models are from $\dagger$~\citet{RuschRus2025} and $\ddagger$~\citet{Farsang2025}. Only the LrcSSM value from \citet{Farsang2025} was performed on the NVIDIA A100.}
\label{tab:tab4}
\scriptsize
\setlength{\tabcolsep}{6pt}
\begin{tabular}{l r r r r r r}
\toprule
Model                   & Heart  & SCP1  & SCP2  & Ethanol & Motor & Worms \\
\midrule
NRDE$^{\dagger}$        & 9{,}539 & 1{,}014 & 1{,}404 & 2{,}256   & 7{,}616 & 5{,}386 \\
NCDE$^{\dagger}$        & 1{,}177 & 973   & 1{,}251 & 2{,}217   & 3{,}778 & 24{,}595 \\
Log-NCDE$^{\dagger}$    & 826    & 635   & 583   & 2{,}056   & 730   & 1{,}956  \\
LRU$^{\dagger}$         & 8      & 9     & 9     & 16      & 51    & 94      \\
S5$^{\dagger}$          & 11     & 17    & 9     & 9       & 16    & 31      \\
Mamba$^{\dagger}$       & 34     & 7     & 32    & 255     & 35    & 122     \\
S6$^{\dagger}$          & 4      & 3     & 7     & 4       & 34    & 68      \\
LinOSS-IMEX$^{\dagger}$ & 4      & 42    & 55    & 48      & 128   & 37      \\
LinOSS-IM$^{\dagger}$   & 7      & 38    & 22    & 8       & 11    & 90      \\
LrcSSM$^{\ddagger}$ (A100) & 23  & 12    & 15    & 15      & 31    & 33      \\
\midrule
PHCSSM w/ STDP          & 18     & 58    & 73    & 106     & 54    & 141     \\
PHCSSM w/o STDP         & 15     & 48    & 63    & 99      & 43    & 131     \\
\bottomrule
\end{tabular}
\end{table}

\subsection{Ablation Study}
\label{sec:exp-ablation}

To quantify the per-constraint contribution of the five biological mechanisms to accuracy and cross-seed stability, each mechanism is removed in turn while keeping the remaining four active and hyperparameters fixed at the per-dataset selected configuration (Table~\ref{tab:tab5}). The five ablated constraints divide into three groups by behaviour. (i)~Dale's Law dominates: removal causes the largest mean accuracy drop in the comparison (7.4~pp across six datasets), driven primarily by a 21.1~pp drop on EigenWorms and a 10.8~pp drop on EthanolConcentration, alongside dramatic variance amplification on the two longest benchmarks (EigenWorms cross-seed standard deviation $5.8 \to 15.2$, $+162$\%; SelfRegulationSCP2 $6.3 \to 11.9$, $+89$\%). (ii)~ALIF dynamics and within-recurrence lateral connectivity contribute consistently, with mean drops of 2.3~pp and 0.8~pp respectively and no dataset benefiting from removing either. (iii)~Tsodyks--Markram STP and STDP show genuinely dataset-dependent behaviour: removing STP improves accuracy on Heartbeat (4.2~pp), SelfRegulationSCP2 (1.1~pp), and EthanolConcentration (2.7~pp) and hurts on SelfRegulationSCP1 (0.2~pp), MotorImagery (3.6~pp), and EigenWorms (2.8~pp), yielding a near-neutral mean drop of 0.23~pp; removing STDP improves on Heartbeat (2.9~pp), SelfRegulationSCP1 (0.5~pp), SelfRegulationSCP2 (7.4~pp, the largest single drop in the comparison), and EigenWorms (0.6~pp) but hurts on EthanolConcentration (3.1~pp) and MotorImagery (3.2~pp), with a mean drop of 0.85~pp. No single constraint is universally redundant. A clean dependence on sequence length, neuron dimension, or topology choice is absent for STP and STDP.

\begin{table}[!htbp]
\centering
\caption{Ablation study. Test accuracy (mean~$\pm$~std over 5 seeds) under systematic removal of individual bio constraints.}
\label{tab:tab5}
\scriptsize
\setlength{\tabcolsep}{4pt}
\begin{tabular}{l c c c c c c}
\toprule
Variant & Heartbeat & SCP1 & SCP2 & EthanolConc. & Motor-Im. & EigenWorms \\
\midrule
Full bio-constraints                & 73.9$\pm$3.9 & 80.7$\pm$1.6 & 58.3$\pm$6.3 & 31.1$\pm$6.1 & 54.7$\pm$4.5 & 85.0$\pm$5.8 \\
w/o Dale's Law                      & 71.3$\pm$3.1 & 81.3$\pm$2.1 & 54.0$\pm$11.9 & 20.3$\pm$3.0 & 48.4$\pm$4.0 & 63.9$\pm$15.2 \\
LIF w/o adaptive th.   & 71.3$\pm$5.3 & 77.9$\pm$4.9 & 52.6$\pm$7.8  & 30.6$\pm$4.7 & 52.6$\pm$3.9 & 85.0$\pm$4.6 \\
w/o STP                             & 69.7$\pm$10.5 & 80.9$\pm$2.7 & 57.2$\pm$7.6 & 28.4$\pm$1.1 & 58.3$\pm$3.8 & 87.8$\pm$1.5 \\
w/o STDP                          & 71.0$\pm$2.6 & 80.2$\pm$2.4 & 50.9$\pm$8.1  & 34.2$\pm$5.7 & 57.9$\pm$4.3 & 84.4$\pm$6.7 \\
w/o lateral connection  & 73.6$\pm$4.2 & 80.0$\pm$3.0 & 56.8$\pm$5.5  & 30.6$\pm$5.5 & 53.3$\pm$3.4 & 84.4$\pm$7.5 \\
\bottomrule
\end{tabular}
\end{table}

\subsection{Cross-Mode Equivalence and Cross-Backend Agreement}
\label{sec:exp-crossbackend}

PHCSSM is deployed as a sequential RSNN with the same trained weights, without the ANN-to-SNN 
conversion step required by standard rate-coded pipelines; the asymptotic equivalence of the 
parallel-scan training mode and the sequential RSNN deployment mode is established in 
Section~\ref{sec:method-equiv}. Production training and deployment both fix the within-step 
iteration count at $\Nmax = 12$, with the convergence regulariser $\Lconv$
keeping the within-step recurrence contractive throughout training so that the $\Nmax = 12$ 
truncation operates inside the asymptotic regime; the iteration-to-convergence mode that extends 
$K$ up to 768 on the same trained weights is an analytical probe of the asymptotic limit, not a 
deployment configuration. The three modes are compared within a single H100 GPU backend, and the 
RSNN deployment mode is further measured across four hardware backends (Table~\ref{tab:tab6}).

Within the H100 GPU backend, the PHCSSM training mode and the RSNN deployment mode produce identical 
argmax predictions on 1{,}864 of 1{,}880 pooled test samples (99.15\%); raising to the iteration-to-convergence 
mode lifts agreement to 1{,}879 of 1{,}880 (99.95\%). Test accuracy of the two modes differs by no more
 than 0.35 percentage points on every dataset, well within seed-level variance. On the worst-case dataset (EthanolConcentration), the training-mode versus
  RSNN spike-tensor disagreement on the $\Nmax = 12$ unsettled subset further decays by factors of 1{,}489$\times$ 
  (seed 4567) to 6{,}937$\times$ (seed 6789) as $\Nmax$ increases from 12 to the iteration-to-convergence cap of 
  768, with all five seeds showing monotonic convergence and no architectural plateau.

Across the four hardware backends running the RSNN deployment mode (Table~\ref{tab:tab6}), the two scalar-IEEE-754 backends (x86 CPU and Cortex-M4F) produce bit-identical argmax predictions against the x86 CPU reference on all 1{,}880 test samples, while the two vectorised-arithmetic backends register sub-half-percent argmax drift: the H100 GPU CUDA reduction tree differs on 4 of 1{,}880 (0.21\%) and the Cortex-A76 NEON fused-multiply-add differs on 3 of 1{,}880 (0.16\%). The 7 affected samples plus the 16 within-H100 cross-mode disagreers lie within fp32 rounding distance of the decision boundary; flips are deterministic at fixed seed and reflect IEEE-754 finite-precision properties of the trained weights, not a software defect.

\begin{table}[!htbp]
\centering
\caption{Cross-backend bit-exact agreement in the RSNN deployment mode. Each entry reports the number of samples whose argmax prediction is bit-identical to the x86 CPU JAX-CPU Python reference (the host backend against which the chip-class C port is compiled and verified) on the same H100-trained checkpoint, out of the per-dataset sample count $n$. All backends execute the RSNN deployment mode of Section~\ref{sec:method-equiv}.}
\label{tab:tab6}
\scriptsize
\setlength{\tabcolsep}{4pt}
\resizebox{\textwidth}{!}{%
\begin{tabular}{l c c c c c}
\toprule
Dataset & $n$ & x86 CPU (baseline) & Cortex-M4F & Cortex-A76 (NEON) & H100 GPU (CUDA) \\
\midrule
Heartbeat                  & 310    & 310    & 310    & 307    & 310 \\
SCP1                       & 425    & 425    & 425    & 425    & 425 \\
SCP2                       & 285    & 285    & 285    & 285    & 284 \\
EthanolConcentration       & 395    & 395    & 395    & 395    & 392 \\
MotorImagery               & 285    & 285    & 285    & 285    & 285 \\
EigenWorms                 & 180    & 180    & 180    & 180    & 180 \\
\midrule
Total                      & 1{,}880 & 1{,}880 & 1{,}880 (100.00\%) & 1{,}877 (99.84\%) & 1{,}876 (99.79\%) \\
\bottomrule
\end{tabular}}
\end{table}

\subsection{CPU Deployment Speedup}
\label{sec:exp-cpuspeedup}

PHCSSM's two execution modes serve complementary hardware regimes from a single trained checkpoint: the parallel-scan training mode exploits GPU wide parallelism via the log-depth scan over $T$ (Section~\ref{sec:method-impl}), while the RSNN deployment mode collapses the within-step iteration loop into a single per-timestep update sized for chip-class CPUs. On a single-thread CPU host, the RSNN deployment mode is 7.2- to 35.4-fold faster than the training-mode forward pass, with a cross-dataset geometric mean of 20.5-fold (Table~\ref{tab:tab7}).

The four-way decomposition isolates the two execution modes from the two hardware regimes. The PHCSSM training mode is 497-fold faster on the H100 GPU than on the CPU host (geometric mean 0.18~ms versus 89~ms per sample), when the log-depth parallel scan over $T$ maps directly to GPU SIMD lanes; the RSNN deployment mode is 6.5-fold slower on the GPU than on the CPU (geometric mean 28.5~ms versus 4.4~ms per sample), because the sequential per-timestep update incurs kernel-launch overhead and underutilises GPU wide-parallel hardware. This result indicates the GPU is the right hardware for the PHCSSM training mode and the CPU is the right hardware for the RSNN deployment mode.

\begin{table}[!htbp]
\centering
\caption{Per-sample inference latency for the PHCSSM training mode and the RSNN deployment mode on the same H100-trained checkpoints. Both CPU and GPU columns use a uniform 5-seed $\times$ 5-rep protocol with per-seed median latency aggregated as the geometric mean across seeds. CPU columns: single-thread x86 CPU (JAX-CPU JIT). GPU columns: H100 GPU (JAX-CUDA JIT). $T$ denotes the input sequence length.}
\label{tab:tab7}
\scriptsize
\setlength{\tabcolsep}{6pt}
\resizebox{\textwidth}{!}{%
\begin{tabular}{l r r r r r}
\toprule
Dataset                 & $T$    & PHC CPU (ms) & RSNN CPU (ms) & PHC GPU (ms) & RSNN GPU (ms) \\
\midrule
Heartbeat               & 405    & 9.45    & 1.31  & 0.045 & 7.42 \\
SelfRegulationSCP1      & 896    & 59.82   & 1.93  & 0.111 & 12.97 \\
SelfRegulationSCP2      & 1{,}152 & 129.27 & 5.02  & 0.149 & 16.54 \\
EthanolConcentration    & 1{,}751 & 102.55 & 3.06  & 0.207 & 25.09 \\
MotorImagery            & 3{,}000 & 46.32  & 4.28  & 0.123 & 53.13 \\
EigenWorms              & 17{,}984 & 1{,}473.33 & 41.57 & 1.803 & 251.98 \\
\midrule
Geometric mean          & n.a.   & 89.4    & 4.37  & 0.18  & 28.49 \\
\bottomrule
\end{tabular}}
\end{table}

\subsection{Edge Deployment Latency}
\label{sec:exp-edge}

Edge biomedical applications require sequence models that fit microcontroller-class memory budgets and run at physiological sampling rates on battery power. To test whether PHCSSM RSNN can be deployed end-to-end on this hardware class and how it compares against the JAX-measurable SSM baselines, we measure inference on a commodity Cortex-M4F microcontroller. The reference chip is the STM32L412KB (Cortex-M4F up to 80~MHz, 40~KB SRAM, 128~KB Flash, CR2032-compatible); the deployment verdict generalises to other commodity Cortex-M4/M4F MCUs.

PHCSSM RSNN's end-to-end inference latency, per-timestep streaming latency, Flash and SRAM utilisation, per-inference energy, and CR2032-cell yield are reported in Table~\ref{tab:tab8}. Of the five JAX-measurable SSM baselines (LRU, S5, LinOSS, PD-SSM, LrcSSM), PHCSSM RSNN is the only model whose fp32 weight footprint fits the 128~KB Flash budget on all six benchmarks (9.2 to 39.6~KB; Table~\ref{tab:tab9}). 
These figures are reported for unquantised \texttt{fp32} weights; standard \texttt{int8} quantisation would yield an additional $\sim$4$\times$ Flash reduction, placing PHCSSM RSNN deployment well below the budget of even smaller commodity Cortex-M class MCUs
The 40~KB SRAM budget is met by every measurable baseline, so the binding deployment constraint at this microcontroller scale is weight storage rather than runtime memory.

On larger CPU hosts (x86 CPU and Cortex-A76), PHCSSM RSNN remains competitive with the five JAX-measurable baselines across all six datasets. The 4~ms/sample budget set by 250~Hz ECG and 256~Hz EEG sampling rates is met with at least two orders of magnitude of headroom on the Cortex-A76, indicating physiological-rate streaming is not the binding deployment constraint.

\begin{table}[!htbp]
\centering
\caption{STM32L412KB end-to-end RSNN-mode deployment profile across the six UEA benchmarks. Latencies measured on a NUCLEO-L412KB development board (Cortex-M4F at 16~MHz HSI clock, $V_{DD}=3.3$~V) under the STM32 Arduino-core build; Flash and Static-RAM read from the linker map. Active-mode power is measured at 9.37~mW ($2.84$~mA~$\times$~$3.3$~V via an AD3-supplied 3.3~V rail with multimeter in the DC mA path, USB unplugged); energy per inference is power times per-inference latency. Inferences per CR2032 cell assume 660~mWh.}
\label{tab:tab8}
\scriptsize
\setlength{\tabcolsep}{3pt}
\resizebox{\textwidth}{!}{%
\begin{tabular}{l r r r r r r r r r}
\toprule
Dataset & $T$ & Per-step ($\mu$s) & Per-inf. (ms) & Flash (KB) & Flash \% & SRAM (KB) & SRAM \% & Energy (mJ) & Inf./CR2032 \\
\midrule
Heartbeat               & 405     & 487    & 197      & 26.6 & 20.8 & 1.73 & 4.3 & 1.85   & 1.28M \\
SCP1                    & 896     & 1{,}986 & 1{,}779 & 40.8 & 31.9 & 3.50 & 8.7 & 16.41  & 145k \\
SCP2                    & 1{,}152 & 1{,}898 & 2{,}187 & 41.0 & 32.1 & 3.50 & 8.7 & 20.46  & 116k \\
EthanolConcentration    & 1{,}751 & 1{,}922 & 3{,}365 & 40.2 & 31.4 & 3.50 & 8.7 & 31.41  & 75.6k \\
MotorImagery            & 3{,}000 & 490    & 1{,}471  & 26.8 & 20.9 & 1.73 & 4.3 & 13.57  & 175k \\
EigenWorms              & 17{,}984 & 1{,}947 & 35{,}019 & 41.1 & 32.1 & 3.50 & 8.7 & 326.31 & 7.3k \\
\bottomrule
\end{tabular}}
\end{table}

\begin{table}[!htbp]
\centering
\caption{STM32L412KB deployment feasibility (40~KB SRAM, 128~KB Flash) for PHCSSM RSNN mode and five JAX-measurable SSM baselines at the per-dataset hyperparameters cited in Table~\ref{tab:tab1}. Each cell reports the serialised fp32 weight Flash (KB; M = MB) measured via \texttt{equinox.tree\_serialise\_leaves}. The `Fits into' column reports the smallest commodity ARM Cortex-M MCU Flash class (powers of 2) that holds the worst-case dataset configuration for each model; PHCSSM RSNN is the only model fitting the 128~KB chip-class budget. n.a.~$=$~not applicable (missing per-dataset hyperparameters). C runtime code overhead ($\sim$20~KB on the STM32L412KB, dataset-invariant) is excluded.}
\label{tab:tab9}
\scriptsize
\setlength{\tabcolsep}{5pt}
\begin{tabular}{l r r r r r r l}
\toprule
Model       & Heart & SCP1 & SCP2 & Ethanol & Motor & Worms & Fits into \\
\midrule
\textbf{PHCSSM RSNN} & 9.2  & 39.3  & 39.6   & 38.5  & 9.3   & 39.6 & \textbf{128~KB} \\
S5          & 105   & 54.8  & 334    & 43.7  & 1.56M & 789  & 2~MB \\
LrcSSM      & 329   & 101   & 469    & 320   & 36.6  & 165  & 512~KB \\
LinOSS      & 54.8  & 3.79M & 1.72M  & 34.6  & 363   & 529  & 4~MB \\
PD-SSM      & 90.3  & 132   & n.a.   & 5.31M & n.a.  & 922  & 8~MB \\
LRU         & 3.83M & 1.73M & 1.54M  & 335   & 1.74M & 997  & 4~MB \\
\bottomrule
\end{tabular}
\end{table}

\section{Discussion}
\label{sec:discussion}

The combination delivered by PHCSSM, parallel-scan $\bigO(\log T)$ training of a biologically-constrained recurrent spiking architecture with end-to-end deployment on chip-class hardware, addresses a structural gap that has shaped the divergent evolution of parallel-scan SSMs and biologically-constrained spiking models.

\subsection{Positioning in the parallel-scan vs biological-fidelity landscape}
\label{sec:disc-positioning}

PHCSSM closes a gap between two architectural directions that have been mutually exclusive in prior work. Parallel-scan SSMs (S4, Mamba, LinOSS, LRU, S5) achieved $\bigO(\log T)$ training by enforcing diagonal state transitions that decouple neurons within a timestep, foreclosing the lateral and feedback connectivity, sign-restricted Dale, and state-dependent transmission primitives that biological circuits are built from. Biologically-constrained spiking models preserved these mechanisms but trained sequentially with surrogate-gradient BPTT, which becomes numerically unstable at depth \citep{ZenkeGanguli2018} and has historically confined SNN benchmarks to thousand-timestep tasks; the 17{,}984-timestep EigenWorms benchmark sits an order of magnitude beyond that historical SNN frontier. Three adjacent architectures clarify why preserving the full set of five mechanisms matters in the present configuration. Liquid State Machines \citep{Maass2002} keep the recurrent spiking reservoir but freeze it and train only a linear readout, structurally excluding recurrent-layer plasticity. Recent spiking-SSM hybrids \citep{StanRhodes2024,Zhong2024,Shen2025} recover gradient-trainable recurrent connectivity but drop the non-linear biological primitives that distinguish cortical computation from rate codes. Surrogate-gradient training \citep{ZenkeGanguli2018,Bellec2018} preserves mechanisms and trainability simultaneously, but at sequential $\bigO(T)$ cost. PHCSSM is the first architecture to combine all five biological constraints with parallel-scan $\bigO(\log T)$ training of the recurrent layer.

\subsection{Cross-backend reproducibility and the conventional digital substrate}
\label{sec:disc-repro}

The cross-backend reproducibility pattern (Section~\ref{sec:exp-crossbackend}, Table~\ref{tab:tab6}) admits a structural interpretation that is not specific to PHCSSM. The two scalar-IEEE-754 backends (x86 CPU and Cortex-M4F) produce bit-identical argmax predictions on every test sample; the two vectorised-arithmetic backends (H100 GPU CUDA and Cortex-A76 NEON) each register sub-half-percent argmax drift, deterministic at fixed seed and concentrated on samples within fp32 rounding distance of the decision boundary. The drift reflects the CUDA reduction-tree topology \citep{WhiteheadFitFlorea2011} and the ARM NEON fused-multiply-add semantics. The simpler hardware does not produce worse reproducibility, it produces stricter reproducibility, because scalar IEEE-754 with a fixed reduction order is the most pessimistic numerical contract. The implication for deployment of biologically-constrained spiking models on heterogeneous edge hardware is that the chip-class backend is the reproducibility floor, not the reproducibility ceiling, of the cross-backend chain. The chip-class deployability is itself a paradigm point: PHCSSM runs end-to-end within a coin-cell energy envelope on a commodity Cortex-M4F microcontroller ($\sim$1.85~mJ per Heartbeat inference, Section~\ref{sec:exp-edge}) without recourse to a neuromorphic substrate. Biological mechanism preservation is therefore not a deployment burden on commodity digital hardware; neuromorphic substrates would add further per-spike efficiency but are not a precondition for embedded deployment of biologically-grounded sequence models.

\subsection{Limitations}
\label{sec:disc-limitations}

Beyond the structural choices already discussed, four scope limitations bound the conclusions drawn here. (i)~The evaluation is restricted to binary and multi-class classification at state dimensions of at most 64; regression, generation, larger configurations (4R128, 6R64), and mainstream long-sequence benchmarks such as the Long Range Arena remain outside the scope of this work. (ii)~The reported energy figures (Section~\ref{sec:exp-edge}) are measured at 16~MHz HSI clock on a NUCLEO-L412KB development board with the multimeter in the $V_{DD}$ path; this total includes onboard regulator and LED overhead that bare-chip wearable PCBs would eliminate, and the active-inference current is within the 10~$\mu$A noise floor of the multimeter, so the MCU-only inference-power increment is not isolated by the present instrumentation. (iii)~The PHC framework is instantiated here only as a spiking SSM (PHCSSM); the framework itself is agnostic to the underlying diagonal state-space core, and instantiations with continuous-valued cores (e.g., S5, LinOSS, Mamba) inside the hierarchical-connectome scaffold are not explored. (iv)~The neuron and synapse primitives (ALIF, Tsodyks--Markram STP) and the single excitatory / single inhibitory population structure under Dale's Law are tractable closed-form approximations of cortical dynamics; richer biological ingredients (e.g., compartmental dendrites, multiple ionotropic and neuromodulatory receptor systems, and cortical interneuron subtype heterogeneity such as PV+, SST+, and VIP+ classes) are not represented, and incorporating them would require new parallel-scan formulations and a richer connectivity scaffold than the present per-neuron sign restriction.

\subsection{Future Directions}
\label{sec:disc-future}

Six research extensions follow naturally from the present work. (i)~Direct measurement of the bio-energy-efficiency contribution on a neuromorphic substrate. The chip-class inference-energy figures (Section~\ref{sec:exp-edge}) quantify the full PHCSSM forward pass on a conventional von-Neumann substrate but do not isolate the energy contribution of the biological constraints themselves, since the conventional MCU pays the full multiply-accumulate cost of every spike regardless of value. Whether biological mechanism preservation, which carries no inference-energy advantage on conventional digital hardware, carries an isolable advantage on a neuromorphic substrate that exploits binary spike sparsity at the instruction level remains open, and is conditional on a future neuromorphic substrate supporting the ALIF and STP primitives PHCSSM relies on. A complementary measurement campaign at 80~MHz HSI+PLL on a bare-chip PCB (rather than the NUCLEO development board) would isolate the bare-MCU active-vs-idle current delta currently masked by NUCLEO board overhead, and would establish the realistic energy-per-inference operating point for wearable deployment; the predicted reduction from 16~MHz NUCLEO to 80~MHz bare-chip is approximately 1.4--1.8$\times$. (ii)~Digital twins from empirical connectivity. The hand-designed two-region hierarchical topology mask can be replaced with empirical laminar connectivity derived from tract-tracing atlases (e.g., the Allen Mouse Brain Connectivity Atlas) or whole-brain connectomes (e.g., the FlyWire Drosophila connectome), advancing PHCSSM toward data-driven functional digital twins of specific neural circuits; the bio-prior preservation that this work establishes (ALIF, STP, Dale's Law, STDP) provides the local substrate, with parallel-scan trainability enabling fits to long behavioural recordings that surrogate-gradient SNNs cannot accommodate. (iii)~Plasticity-rule exploration. The current STDP module implements the two-factor (pre, post) Hebbian rule; the effect of different LTP and LTD window shapes and eligibility-trace decay constants on PHCSSM, as well as extensions to three-factor variants that incorporate neuromodulatory signals, have not been characterised. (iv)~STDP at deployment time. The current STDP module is exercised only during offline training. Closed-loop deployment-time on-chip adaptation requires a neuromorphic substrate with local learning support (e.g., Loihi's on-chip learning engine, \citealt{Davies2018}); porting PHCSSM's STDP forward and update path onto such a substrate would enable per-subject signal-drift adaptation in situ without an offline retraining loop. (v)~Closed-loop integration with ECG and EEG sensors, motor-imagery brain-computer interfaces, and neuromodulation devices would exercise the chip-class deployability under in-the-loop signal drift. (vi)~Wider benchmark evaluation. The six-dataset UEA-MTSCA suite is suggestive of dataset-dependent biological-mechanism utility patterns but is too narrow for confident generalisation; a systematically wider suite (regression and generation tasks, longer-context Long Range Arena tasks) is required to firm the per-mechanism interpretation.
(vii)~Quantisation-aware compression. The deployment-mode Flash and SRAM figures reported here (Section~\ref{sec:exp-edge}, Table~\ref{tab:tab9}) are for unquantised \texttt{fp32} weights; standard post-training \texttt{int8} or sub-byte quantisation-aware training would compress the Flash footprint by a further $\sim$4$\times$ to $\sim$8$\times$ while remaining within the cross-backend bit-error envelope characterised in Section~\ref{sec:exp-crossbackend}. This would bring PHCSSM RSNN within the Flash budget of Cortex-M0/M3 class MCUs an order of magnitude smaller than the STM32L412KB reference chip used here, extending the chip-class deployability frontier downward to even more constrained edge devices.

\section{Conclusion}
\label{sec:conclusion}

PHCSSM is a spiking state-space model integrating five biological constraints (ALIF dynamics, short-term plasticity, Dale's Law with E/I-asymmetric topology, hierarchical connectome topology, and STDP) within a fully parallelizable $\bigO(\log T)$ framework, natively trained as a recurrent spiking network without ANN-to-SNN conversion. On the six UEA-MTSCA physiological benchmarks PHCSSM is competitive across sequence lengths from 405 to 17{,}984 timesteps at 1{,}312 to 4{,}891 trainable parameters, between one and three orders of magnitude smaller than every comparison baseline, while uniquely integrating all five biological mechanisms; per-mechanism ablation isolates Dale's Law as the dominant accuracy and variance regulariser.

The same trained weights deploy without retraining as a sequential RSNN whose asymptotic equivalence to the parallel-scan training mode is established. Cross-backend verification across H100 GPU, x86 CPU, Cortex-A76, and Cortex-M4F yields bit-identical predictions on the two scalar-IEEE-754 backends and sub-half-percent argmax drift on the two vectorised-arithmetic backends. On a single-thread CPU host the RSNN deployment mode runs 7- to 35-fold faster than the training-mode forward pass on the same trained checkpoint, and end-to-end deployment on the Cortex-M4F (40~KB SRAM, 128~KB Flash) is verified within a coin-cell energy envelope. These results establish that biologically grounded structural priors function as enablers, rather than burdens, of parameter-efficient and edge-deployable sequence modelling.

\section*{Data and Code Availability}
The source code will be made publicly available upon acceptance of this manuscript, subject to applicable intellectual property restrictions. The datasets used in this study are publicly available from the UEA Multivariate Time-Series Classification Archive.

\section*{Acknowledgements}
This work was supported by the National Science and Technology Council (NSTC), Taiwan (NSTC 114-2320-B-A49-027-; NSTC 114-2634-F-A49-006-; NSTC 114-2321-B-A49-003-; NSTC 114-2321-B-A49-014-).

\section*{CRediT authorship contribution statement}
\textbf{P.-H. Chiang}: Conceptualization, Data curation, Formal analysis, Funding acquisition, Investigation, Methodology, Project administration, Resources, Software, Validation, Visualization, Writing -- original draft, Writing -- review and editing.

\section*{Declaration of competing interest}
The author declares the following competing interests: a patent application related to the work described in this paper has been filed.

\section*{Declaration of generative AI and AI-assisted technologies in the manuscript preparation process}
During the preparation of this work, the author used Claude (Anthropic) and Gemini for assistance with manuscript editing and code development. The author reviewed and verified all outputs and takes full responsibility for the content of this work.


\end{document}